\documentclass[prb,preprint]{revtex4}

\usepackage{amsmath}
\usepackage{graphicx}

\newcommand{\bra}[1]{\langle#1|}

\newcommand{\ket}[1]{|#1\rangle}

\newcommand{\cfg} [1] {\{ #1 \}}
\graphicspath{{converted_graphics/}}
\begin{document}
% as submitted to arXiv cond-mat and Phys Rev E, Sept. 3, 2009, as of 9:30 a.m. EDT

\title{Exact energy spectrum of a two-temperature kinetic Ising model}
\author{I. Mazilu}
\email{mazilui@wlu.edu}
\author{H. T. Williams}
\email{williamsh@wlu.edu}
\affiliation{Department of Physics and Engineering, Washington and Lee University, Lexington, Virginia 24450, USA}
\date{September 3, 2009}
\begin{abstract}
The exact energy spectrum is developed for a two temperature kinetic Ising spin chain, and its dual reaction diffusion system with spatially alternating pair annihilation and creation rates. Symmetries of the system pseudo-Hamiltonian that enable calculation of the spectrum are also used to derive explicit state vectors for small system sizes, and to make observations regarding state vectors in the general case. 
Physical consequences of the surprisingly simple form for the eigenvalues are also discussed.
\end{abstract}
\maketitle

\section{Introduction}

Over the last three decades, an increasing number of condensed matter theorists are devoting their efforts to understanding complex collective behavior of far-from-equilibrium systems using methods that range from easily accessible computer simulations to sophisticated theoretical studies. Although great progress has been made, a comprehensive theoretical framework is still lacking. In this context, low-dimensional systems are of particular interest since their simplicity permits analytical and numerical solutions \cite {privman,ziabook,others} and these non-trivial solutions shed light on related, more complicated higher-dimensional models. \newline

This paper presents the exact energy spectrum of two closely related one-dimensional non-equilibrium models: a kinetic Ising chain (KISC) with cells coupled alternately to one of two temperature baths, with generalized Glauber dynamics \cite {glauber}, and its dual counterpart, a reaction diffusion system (RDS) with spatially alternating pair annihilation and creation rates. Interest in these models is also motivated by their experimental applications. Multi-temperature spin systems are fairly common: nuclear magnetic resonance in an external magnetic field is an example; a lattice of nuclei in a solid prepared at a finite spin temperature \cite{schmuser} is another. On the other hand, the RDS model with spatially alternating annihilation and creation rates is known to  describe the dynamics of photo-excited solitons in polymers \cite{kuroda}. Mobilia \it et al. \rm  proposed an experimental realization of the RDS model with alternating rates in MX chain compounds using a laser with spatially modulated power output \cite{mobilia}.\newline

The two-temperature kinetic Ising model (KISC), was first introduced by Racz and Zia \cite {RZ(kinetic)} who calculated exactly the two-point correlation functions for the steady state. Using a perturbation expansion of the master equation, Schmuser and Schmittmann \cite{schmuser} calculated the first two corrections to the equilibrium Boltzmann distribution. Mobilia \it et al.\rm \cite{mobilia} found an analytical solution for the full dynamics (magnetization, particle density, and all correlation functions) of this non-equilibrium spin chain and its related reaction-diffusion model using a generating function approach. Outstanding challenges include knowledge of the exact energy spectrum of these models, and a compact expression for their steady states. 

Our study brings us one step closer to achieving this goal. Using the standard mapping \cite{henkel} of reaction diffusion models onto integrable quantum chains, the RDS model can be expressed in a ``free fermion" form, by defining a quadratic non-Hermitian ``stochastic Hamiltonian"\cite{mobilia}. This operator can be diagonalized as long as certain constraints are obeyed \cite {grynberg}. In this paper, we derive the exact energy spectrum of this pseudo-Hamiltonian. We also utilize symmetry considerations to extract the energy eigenvalues and associated eigenstates for some small system sizes with the goal of finding a general pattern for the steady states of these models. 

Our paper is organized as follows: In Section 1, we give an overview of the models. Next (Section 2), we describe the symmetries exhibited by the pseudo-Hamiltonian operator, and their role in the diagonalization process. We present exact solutions (eigenvalues and eigenvectors) for some small system sizes in Section 3. Following some standard technical steps (Jordan Wigner transformation, discrete Fourier transform and a generalized Bogoliubov transformation) we derive closed-form expressions for the eigenvalues, and a methodology for extracting the eigenvectors. Section 5 presents a summary of our results and some possible generalizations for these models. \newline. 

\section{Overview of model}

Two equivalent one-dimensional models motivate the work herein: the kinetic Ising spin chain (KISC) and its associated reaction-diffusion model (RDS.) The KISC model parallels the one-dimensional Ising model.  We postulate a lattice of $N$ side-by-side cells, numbered $n = 1, 2, \ldots, N$, arranged in a ring such that cell $n=N$ is considered adjacent to cell $n=1$.  $N$ is restricted to even values.  Each cell has a single degree of freedom with two possible values:  $-1$, that can be thought a cell occupied by a particle with spin down; and $+1$, describing a cell occupied by a particle with spin up. Each cell interacts with its two nearest neighbors, as well as being in contact with a heat bath at one of two temperatures --- $T_{e}$ for even-numbered cells and $T_{o}$ for odd numbered cells.  If $T_{e}\neq T_{o}$ the system cannot achieve equilibrium: each heat bath tries to drive the system towards a different equilibrium state. As a result, energy flows continuously between the even cell sublattice and the odd. Configuration $C$ (a list of the states of the $N$ cells) changes into a different configuration $C' $ with generalized Glauber transition rates $r\left[ C \rightarrow C' \right] $.  Rate $r$ is non-zero only if $C$ and $C'$ differ only in the spin of a single particle.  The rate at which site $n$ has its spin flipped is given by:
\begin{equation} r_n = \frac{1}{2} - \frac{\gamma_n}{4}d_{n}(d_{n-1}+d_{n+1}) \label{rate} \end{equation}
where $k_B$ is Boltzmann's constant, the factor $\gamma_n$ ($0 \leq \gamma_n \leq 1$) is related to the temperature of cell $n$ by
\begin{equation}
\gamma _{n}=\left\{ 
\begin{array}{c}
\tanh (\frac{2}{k_{B}T_{e}}) \\ 
\\ 
\tanh (\frac{2}{k_{B}T_{o}})
\end{array}
\right. \quad \mbox{for\quad }
\begin{array}{c}
n \mbox{ even} \\ 
\\ 
n \mbox{ odd} ,
\end{array}
\end{equation}
and $d_n$ is the state ($+1$ or $-1$) of the $n$-th cell.  This rate equation prescribes a spin flip rate for a cell of $1/2$ if cells to the left and right have opposite spins, $(1-\gamma_n)/2$ if adjacent spins are the same and the same as that of cell $n$, and $(1+\gamma_n)/2$ if adjacent spins are the same and opposite that of cell $n$.  The time scale is arbitrary.

The KISC model is mapped onto an equivalent reaction-diffusion model with spatially alternating pair creation and annihilation rates in the following way.  A dual lattice of $N$ sites is established, in which a site in the dual lattice is associted with the boundary between two sites in the KISC lattice. A pair of adjacent KISC spins with opposite signs is identified with a particle in the dual lattice; adjacent spins with the same sign is identified with the absence of a particle (a hole.) A spin flip in the KISC model translates into either diffusion of particles on the dual lattice with equal left-right rates, or pair creation or annihilation with different rates.  Transition rates between configurations in the KISC system become diffusion, pair creation and annihilation rates in the RDS system, as shown in Table 1.

Time evolution of these systems is described by the master equation, expressing conservation of probability assuming a continuous-time dynamics. The probability $P(C ,t)$ of finding the system in configuration $C $ at time $t$ increases due to transfer of probability into $C$ from other configurations, and decreases as $C$ passes probability into others,  in such a way that $\sum_{C }P(C ,t)=1$ for all $t$.  The evolution of probability $P(C ,t)$  is described by transition rates $r\left[ C \rightarrow C' \right]$, the probability per unit time that configuration $C$ changes into a different configuration $C' $.  The master equation is:
\begin{equation} 
\frac{dP(C ,t)}{dt}=\sum_{C' \neq C}\left\{ r\left[ C' \rightarrow C \right] P(C',t)
 -r\left[ C \rightarrow C'\right] P(C ,t)\right\}   \label{master}
\end{equation}
in which the first term on the right represents the gain in probability of configuration $C$ due to transitions from other configurations, and the second represents losses due to $C$ transforming into other configurations. 

We utilize Dirac notation to represent each configuration as $\ket{C}$.  From this we build a vector representation of a probabilistic superposition of all possible configurations of a system:
\begin{equation} 
\ket{P(t)}=\sum_{C}P(C,t)\ket{C} .
\end{equation}
The master equation can now be re-expressed in terms of this vector as:
\begin{equation} 
\frac{d}{dt} \ket{P(t)} = -H |P(t)\rangle \label{me1}
\end{equation}
where the pseudo-Hamiltonian $H$ is a $2^N \times 2^N$ matrix, with matrix elements 
\begin{eqnarray}
  \bra{C'}H\ket{C}  &=& -r(C \rightarrow C'), C'\neq C \\
  \bra{C}H\ket{C} &=& \sum_{C' \neq C}r(C \rightarrow C') .
\end{eqnarray}
A formal solution to Eq. \ref{me1} can be written as $\ket{P(t)} = e^{-Ht} \ket{P(0)}$. 
Our goal is to investigate the eigenvalues of operator $H$ in order to explain the system's time dependence. 

From this point, we shall focus on the RDS model. We follow the precedent of representing the ``particles" and ``holes" in the dual lattice by a spin one-half model:  a particle is represented by spin up (and thus $\ket{1}$, a hole becomes a spin down ($\ket{0}$.)  From the above formalism comes the definition of the probability-conserving operator $H$ that controls the system's time dependence:
%\pagebreak
\begin{eqnarray}
-2H&=&\sum_{j even}[\sigma^{+}_{j}\sigma^{-}_{j+1}+\sigma^{-}_{j}\sigma^{+}_{j+1}+(1+\gamma_{e})\sigma^{+}_{j}\sigma^{+}_{j+1}+
(1-\gamma_{e})\sigma^{-}_{j}\sigma^{-}_{j+1}-\gamma_{e}(\sigma^{-}_{j}\sigma^{+}_{j}+
 \sigma^{-}_{j+1}\sigma^{+}_{j+1}) \nonumber \\
&-&(1-\gamma_{e})]  + \sum_{j odd}[\sigma^{+}_{j}\sigma^{-}_{j+1}+\sigma^{-}_{j}\sigma^{+}_{j+1}+(1+\gamma_{o})\sigma^{+}_{j}\sigma^{+}_{j+1}+ (1-\gamma_{o})\sigma^{-}_{j}\sigma^{-}_{j+1} \nonumber \\
&-&\gamma_{o}(\sigma^{-}_{j}\sigma^{+}_{j}+ \sigma^{-}_{j+1}\sigma^{+}_{j+1})-(1-\gamma_{o})]. \label{Hop}
\end{eqnarray}
The operators $\sigma_n^{+}$ and $\sigma_n^{-}$ are the Pauli spin raising and lowering operators on the $n$-th cell:
\[ \sigma_n^{+} \ket{0}_n = \ket{1}_n , \,\,\,  \sigma_n^{+} \ket{1}_n = 0 , \,\,\, \sigma_n^{-} \ket{0}_n = 0, \,\,\, \mbox{and} \,\,\, \sigma_n^{-} \ket{1}_n = \ket{0}_n . \]

It has been shown that the eigenvalues and eigenvectors of $H$ can be found if the ``free fermion constraint" is obeyed \cite{grynberg}. For the RDS model, this means that the sum of local diffusion rates is equal to the sum of local pair creation and annihilation rates (in our case $ \frac{1}{2}+\frac{1}{2}=\frac{1+\gamma_{o,e}}{2}+\frac{1-\gamma_{o,e}}{2} $.) This constraint assures the biliniarity of the $H$ operator, and, consequently, an exact solution for the problem.

\section{Symmetries of the $H$ operator}
 Symmetries exhibited by the $H$ operator, Eq. \ref{Hop}, affect the form of its eigenvalues and eigenvectors, and in some cases aid in the process of determining them.  In this section our goal is to exhibit these symmetries towards the goal of direct calculation of eigenvalues and eigenvectors of $H$ for small values of $N$.

Because the fundamental process described by the rate equation, Eq. \ref{rate}, corresponds to the simultaneous flipping of two spins in the RDS model, the $H$ operator does not change the 'spin-parity' of a state, i.e. states with an even number of up spins are transformed by $H$ into states of only even number of up spins, and likewise for odd numbers of up spins.  This symmetry immediately separates the configuration space into two sub-spaces of the same dimensionality ($2^N/2$) which do not interact.  Thus the $2^N \times 2^N$ $H$ matrix is reduced to two equal-sized diagonal blocks by proper ordering of the configuration basis states.

The $H$ matrix is also invariant to a translation of the ring of cells by an even number of cells to the right or left. Thus $H$ commutes with the operator that invokes this translation, and simultaneous eigenstates of the two operators can be found.  Such eigenstates are conveniently written as sums of the form
\begin{equation} \ket{\mathcal{C}_q} =  \sum_{n=0}^{N/2-1} e^{i n q \frac{4 \pi}{N}} \, \ket{C_n} , \label{ft} \end{equation}
($q=0, \ldots , N/2-1$), 
where $\ket{C_0}$ is a spin configuration (e.g. $\ket{011000}$ representing six spins, the second and third ones up, for the $1 \times 6$ case) and $\ket{C_n}$ is the same configuration pushed $2n$ cells to the right using periodic boundary conditions (thus if $\ket{C_0} = \ket{011000}$, then $\ket{C_2} = \ket{100001}$.)  The translation symmetry of the pseudo-Hamiltonian implies that $H$ does not mix states of the form $\ket{\mathcal{C}_q }$ which have different $q$ values.  Thus within each of the two major sub-blocks of the $H$ matrix (one of even spin parity, one of odd spin parity) there are $N/2$ smaller sub-blocks, each with a different value of $q$.  This symmetry has enabled relatively straightforward extraction of eigenvalues and eigenstates for even $N$ values up to $N=8$ (with a $256$-dimensional configuration space.)

Two additional symmetries are apparent from the form of $H$ given above, that provide further information regarding the form of the eigenvalues and eigenvectors.  

The simultaneous translation of the spin chain by a single site (cell $n$ becomes cell $n+1$) along with the interchage of values $\gamma_e \leftrightarrow \gamma_o$ leaves $H$ invariant.  If we use $\mathcal{X}$ to represent this transformation, it follows that $\mathcal{X}^2 = I$.  If $\ket{\psi}$ is an eigenstate of $H$ with eigenvalue $E$, then
\[ \mathcal{X} H  \ket{\psi} = H \mathcal{X} \ket{\psi} = \mathcal{X} E \mathcal{X}^{-1} \mathcal{X} \ket{\psi} . \]
This leads to several possibilities:
\begin{itemize}
   \item $E$ is invariant under the interchange $\gamma_e \leftrightarrow \gamma_o$ and either $\mathcal{X} \ket{\psi}$ is a constant multiple of $ \ket{\psi}$, or $\mathcal{X} \ket{\psi}$ produces another eigenstate of $H$ distinct from $ \ket{\psi}$ but with the same eigenvalue.
   \item $E$ is not invariant under the interchange $\gamma_e \leftrightarrow \gamma_o$, but instead transforms to another distinct eigenvalue of $H$, and $\mathcal{X} \ket{\psi}$ becomes a corresponding eigenstate.
\end{itemize}
Explicit diagonalization for small $N$, shown in the next section, suggests that the eigenvalues are invariant under $\gamma_e \leftrightarrow \gamma_o$.  The general solution for the eigenvalues presented in a later section show this to be true for all even values of $N$.

Another symmetry operation leaving $H$ invariant consists in changing the sign of both $\gamma$ constants and simultaneously flipping every spin.  If we use $\mathcal{X}_A$ to represent this transformation, algebra similar to that of the prior paragraph leads to the following:
\begin{itemize}
   \item $E$ is invariant under the change in sign of both $\gamma$'s  and either $\mathcal{X_A} \ket{\psi}$ is a constant multiple of $ \ket{\psi}$, or $\mathcal{X_A} \ket{\psi}$ produces another eigenstate of $H$ distinct from $ \ket{\psi}$ but with the same eigenvalue.
   \item $E$ is not invariant under the change in sign of both $\gamma$'s, but instead transforms to another distinct eigenvalue of $H$, and $\mathcal{X_A} \ket{\psi}$ becomes a corresponding eigenstate.
\end{itemize}
Explicit diagonalization for small $N$ (following section) suggests also that the eigenvalues are invariant under the change in sign of both $\gamma$'s. This, too, is later proven true for all $N$.

\section{Explicit diagonalization of $H$ \label{sect2}}

The number of configurations for a given value of $N$, thus the dimensionality of the square matrix $H$, is $2^N$.  This prohibits explicit diagonalization of $H$ for all but the smallest values of $N$.  The symmetries of $H$ enable straightforward use of computer algebra software to find eigenvectors and eigenvalues for several cases.  In this section some explicit results for $N = 2, 4, 6, 8$, so calculated, are exhibited and discussed.

\subsection{$N=2$ case}
The simplest case, $N=2$, while simple is not quite trivial, and its solutions are useful for establishing patterns for eigenvalues and eigenvectors of $H$.  For this case, there are four basic configurations ($\ket{00}, \ket{11}, \ket{01},\ket{10}$) and thus the $H$ matrix is $4 \times 4$.  The operation of shifting each configuration two cells to the right is equivalent to the identity operation, and this plays no useful role for this case.  Spin parity does play a role, indicating that $H$ simplifies into two $2 \times 2$ diagonal blocks if the basis configurations are ordered as above (even spin-parity states $\ket{00}$ and $\ket{11}$ form a basis for the even subspace, odd spin-parity states $\ket{01}$ and $\ket{10}$ form a basis for the odd subspace.) The $H$ matrix for this basis is:

\[  H = \left( \begin{array}{cccc} 
		1+\frac{\gamma_e + \gamma_o}2 & -1 +  \frac{\gamma_e + \gamma_o}2 & 0 & 0 \\
		-1-\frac{\gamma_e + \gamma_o}2 & 1-\frac{\gamma_e + \gamma_o}2 & 0 & 0 \\
		0 & 0 & 1 & -1 \\
		0 & 0 & -1 & 1
    \end{array}  \right) .\]
The upper $2 \times 2$ block has two eigenvalues:  $0$, the ground state with  eigenvector 
\[   (2- \gamma_e -\gamma_o) \cfg{00} + (2+ \gamma_e +\gamma_o) \cfg{11}  ; \]
and eigenvalue $2$ with eigenvector 
\[  \cfg{00} - \cfg{11}  . \]
The lower block also has two:  eigenvalue $0$ with  eigenvector 
\[   \cfg{01} + \cfg{10}  ; \]
and eigenvalue $2$ with normalized eigenvector 
\[   \cfg{01} - \cfg{10}  . \]

The construction of $H$ guarantees that its eigenvalues are non-negative, and that its lowest eigenvalue is zero.  These results show a twofold degeneracy of the zero eigenvalue, with one eigenstate in the even subspace, and the other in the odd subspace.  This happens as well in the following special cases presented below, and is a general property for all $N$.  

\subsection{$N = 4$ case}

This case requires the diagonalization of a $16 \times 16$ matrix, so utilization of symmetries of $H$ is useful, if not essential.  Both the spin parity symmetry and the translation symmetry play a role in this case.  $H$ breaks into two $8 \times 8$ diagonal blocks immediately - one of even spin parity states, the other of odd.  Within each block, there are two smaller blocks when the basis states are transformed as in Eq. \ref{ft}.  We characterize basis states in the notation of Eq. \ref{ft} by their $q$-values.  Of the eight even spin-parity states, six are $q=0$ states, 
\[ \left( \begin{array}{c} \ket{0000} \\ \ket{1111} \\ \ket{0101} \\ \ket{1010} \\ \ket{1100}+\ket{0011} \\
 \ket{1001}+\ket{0110}  \end{array}    \right),  \]
and the remaining two are $q=1$ states,
\[ \left( \begin{array}{c} \ket{1001}-\ket{0110} \\ \ket{1100}-\ket{0011} \end{array}    \right) . \]
Each of these separately are bases for a diagonal block of $H$, one $6 \times 6$, another $2 \times 2$.

The $8 \times 8$ even-parity subspace block for $H$, using these eight basis states, $q=0$ states first, is

\[  \left(  \begin{array}{cccccccc}
      2+\gamma_e + \gamma_o &0 &0 &0 & \frac{-1+\gamma_e}{2} &\frac{-1+\gamma_o}{2} &0 &0  \\
      0 & 2-\gamma_e - \gamma_o &0 &0 &\frac{-1-\gamma_e}{2} &\frac{-1-\gamma_o}{2} &0 &0  \\
      0 &0 &2 &0 &-\frac12 &-\frac12 &0 &0 \\
      0 &0 &0 &2 &-\frac12 &-\frac12 &0 &0 \\
      -1-\gamma_e &-1+\gamma_e &-1 &-1 &2 &0 &0 &0  \\
       -1-\gamma_o &-1+\gamma_o &-1 &-1 &0 &1 &0 &0 \\
      0&0 &0 &0 &0 &0 &2 &0  \\
      0&0 &0 &0 &0 &0 &0 &2 
      \end{array}  \right) . \]

The upper $6 \times 6$ block ($q = 0$) of this matrix has eigenvalues $0, 2, 2, 4, 2- \sqrt{2 \gamma_e \gamma_o}$, and $2+\sqrt{2 \gamma_e \gamma_o}$.  Corresponding eigenvectors are displayed in Fig. 2.

Of the remaining eight odd spin-parity basis states, four can be cast as $q=0$ states, with basis vectors
\[ \left( \begin{array}{c} \ket{0010}+\ket{1000} \\ \ket{1101}+\ket{0111} \\ \ket{0001}+\ket{0100} \\ \ket{1011}+\ket{1110}  \end{array}    \right) ,  \]
and four as $q=1$ states, with basis vectors
\[ \left( \begin{array}{c} \ket{0010}-\ket{1000} \\ \ket{1101}-\ket{0111} \\ \ket{0001}-\ket{0100} \\ \ket{1011}-\ket{1110}  \end{array}    \right) .  \]
 Using this basis, with the $q=0$ states in the first four positions, this $8 \times 8$ diagonal block of the Hamiltonian from the odd spin parity basis breaks into a $4 \times 4$ block and two $2 \times 2$ blocks:
\[  \left(  \begin{array}{cccccccc}
      
      2+\frac{\gamma_e+\gamma_o}{2} &0 &-1 &-1+\frac{\gamma_e+\gamma_o}{2} &0 &0 &0 &0 \\
      0 &2-\frac{\gamma_e+\gamma_o}{2} &-1-\frac{\gamma_e+\gamma_o}{2} &-1 &0 &0 &0 &0 \\
      -1 &-1 &2+\frac{\gamma_e+\gamma_o}{2} &0 &0 &0 &0 &0 \\
      -1-\frac{\gamma_e+\gamma_o}{2} &-1 &0 &2-\frac{\gamma_e+\gamma_o}{2} &0 &0 &0 &0 \\
      0 &0 &0 &0 &2+\frac{\gamma_e+\gamma_o}{2} &\frac{-\gamma_e+\gamma_o}{2} &0 &0 \\
      0 &0 &0 &0 &\frac{\gamma_e-\gamma_o}{2} &2-\frac{\gamma_e+\gamma_o}{2} &0 &0 \\
      0 &0 &0 &0 &0 &0 &2+\frac{\gamma_e+\gamma_o}{2} &\frac{\gamma_e-\gamma_o}{2} \\
      0 &0 &0 &0 &0 &0 &\frac{-\gamma_e+\gamma_o}{2} & 2-\frac{\gamma_e+\gamma_o}{2}
     \end{array}  \right)  \]

The $4 \times 4$ block has eigenvalues $0, 2, 2$, and $4$; each of the succeeding $2 \times 2$ blocks has eigenvalues $2 \pm \sqrt{\gamma_e \gamma_o}$.  Corresponding eigenvectors are displayed in Fig. 2.

Note that as was the case for $N=2$, there is a twofold degenerate ground state (eigenvalue $0$) with one eigenstate in the even subspace and the other in the odd subspace.  

\subsection{$N = 6$ case}

The configuration space for this case is of dimension $2^6 = 64$.  As suggested by the results of the $N = 4$ case, the form of the eigenvalues is quite complicated, and will not be displayed herein.  The following discussion outlines the steps in solving for eigenvalues and eigenvectors, and exhibits the eigenvalues.

  $H$ breaks into two $32 \times 32$ diagonal blocks  - one of even spin-parity states, the other of odd.  Within each block, there are now three smaller blocks when the basis states are replaced by linear combinations as in Eq. \ref{ft}.  Eigenstates in the respective sub-block bases are of the forms $\ket{ijklmn}+ \ket{klmnij}+\ket{mnijkl}$ (sub-block 1, $q=0$), $\ket{ijklmn}+ e^{i  \frac{2 \pi}{3}}\ket{klmnij}+e^{i  \frac{4 \pi}{3}}\ket{mnijkl}$ (sub-block 2, $q=1$), and $\ket{ijklmn}+ e^{i  \frac{4 \pi}{3}}\ket{klmnij}+e^{i  \frac{2 \pi}{3}}\ket{mnijkl}$ (sub-block 3, $q=2$.)  Note that there are four states, namely $\ket{000000}$, $\ket{010101}$, $\ket{101010}$, and $\ket{111111}$  that transform into themselves under forward translation by two cells, and thus ``interact" (through $H$) only with other states in the $q=0$ sub-block.  The first and fourth of these states are of even spin parity and are  part of a $12 \times 12$ block of even parity states with $q=0$:
\[ \left( \begin{array}{c}
  \ket{000000} \\ \ket{111111} \\ \ket{110000} + \ket{001100}+\ket{000011} \\ \ket{011000}+ \ket{100001}+\ket{011000} \\
  \ket{001111} + \ket{110011}+\ket{111100} \\ \ket{100111}+ \ket{111001}+\ket{011110} \\ \ket{001001}+ \ket{010010}+\ket{100100} \\
  \ket{010001} + \ket{010100}+\ket{000101} \\ \ket{101000}+ \ket{001010}+\ket{100010} \\ \ket{101110}+ \ket{101011}+\ket{111010} \\ 
  \ket{110101} + \ket{011101}+\ket{010111} \\ \ket{110110}+ \ket{101101}+\ket{011011} \end{array}    \right) . \]

Commercially available algebraic software was used to diagonalize this and the other blocks for this case (as for the previous cases.) The twelve eigenvalues for this block are $E = 0$ (the even spin parity ground state), $E = 2$ with degeneracy $3$, $E = 4$ with degeneracy $3$, $E = 6$, $E = 2 \pm \sqrt{3 \gamma_e \gamma_o}$, and $E = 4 \pm \sqrt{3 \gamma_e \gamma_o}$

The remaining two states that transfer into themselves under translation by two sites --- ($\ket{010101}, \ket{101010}$) --- are part of the $q=0$ sub-block for odd spin-parity. The twelve states are
\[ \left( \begin{array}{c}
  \ket{010101} \\ \ket{101010} \\ \ket{010000}+ \ket{000100}+\ket{000001} \\ \ket{100000}+ \ket{001000}+\ket{000010} \\
  \ket{100101} + \ket{011001}+\ket{010110} \\ \ket{101001}+ \ket{0111010}+\ket{100110} \\ \ket{110001}+ \ket{011100}+\ket{000111} \\
 \ket{110100} + \ket{001101}+\ket{010011} \\ \ket{110010}+ \ket{101100}+\ket{001011} \\ \ket{111000}+ \ket{001110}+\ket{100011}, \\ 
  \ket{111110} + \ket{101111}+\ket{111011} \\ \ket{111101}+ \ket{011111}+\ket{110111} \end{array}    \right) , \] 
and the eigenvalues for this block are $E = 0$ (the odd spin parity ground state), $E = 2$ with degeneracy $3$, $E = 4$ with degeneracy $3$, $E = 6$, $E = 2 \pm \sqrt{ \gamma_e \gamma_o}$, and $E = 4 \pm \sqrt{ \gamma_e \gamma_o}$.

There are two remaining even spin-parity sub-blocks, both $10 \times 10$ and both have a basis set of the form
\[ \left( \begin{array}{c}
   \ket{110000} + n \ket{001100}+ n^2 \ket{000011} \\ \ket{011000}+ n \ket{100001}+ n^2 \ket{011000} \\ \ket{001111}+ n \ket{110011}+ n^2 \ket{111100} \\
  \ket{100111}  + n \ket{111001}+ n^2\ket{011110} \\ \ket{001001}+ n \ket{010010}+n^2 \ket{100100} \\ \ket{010001}+ n \ket{010010}+n^2\ket{100100} \\
  \ket{101000}  + n \ket{001010}+n^2 \ket{100010} \\ \ket{101110}+ n \ket{101011}+n^2 \ket{111010} \\  \ket{110101}+ n\ket{011101}+ n^2 \ket{010111} \\  \ket{110110}  + n \ket{101101}+n^2 \ket{011011}  \end{array}    \right) . \]
This is the even spin-parity sub-block for $q=1$ when $n = e^{i 2 \pi / 3}$, and even spin-parity sub-block for $q=2$ when $n = e^{i 4 \pi / 3}$.  For either of these blocks, the family of eigenvalues is the same:  $E = 2$,  $E = 4$,  $E = 2 \pm \sqrt{3 \gamma_e \gamma_o /4}$ each with degeneracy $2$, and $E = 4 \pm \sqrt{3 \gamma_e \gamma_o /4}$ each with degeneracy $2$.

The final two blocks are odd parity blocks, each $10 \times 10$, with basis sets of the form:
\[ \left( \begin{array}{c}
   \ket{010000} + n \ket{000100}+ n^2 \ket{000001} \\ \ket{100000}+ n \ket{001000}+ n^2 \ket{000010} \\ \ket{100101}+ n \ket{011001}+ n^2 \ket{010110}\\
   \ket{101001}  + n \ket{011010}+ n^2\ket{100110} \\ \ket{110001}+ n \ket{011100}+n^2 \ket{000111} \\ \ket{110100}+ n \ket{001101}+n^2\ket{010011} \\
  \ket{110010}  + n \ket{101100}+n^2 \ket{001011} \\ \ket{111000}+ n \ket{001110}+n^2 \ket{100011} \\  \ket{111110}+ n\ket{101111}+ n^2 \ket{111011} \\ 
  \ket{111101}  + n \ket{011111}+n^2 \ket{110111}  \end{array}    \right) . \]
This is the odd spin parity sub-block for $q=1$ when $n = e^{i 2 \pi / 3}$, and odd spin parity sub-block for $q=2$ when $n = e^{i 4 \pi / 3}$.  For either of these blocks, the family of eigenvalues is the same:  $E = 2$,  $E = 4$,  $E = 2 \pm \sqrt{ \gamma_e \gamma_o} /2$, $E = 4 \pm \sqrt{ \gamma_e \gamma_o} /2$ , $E = 2 \pm 3 \sqrt{ \gamma_e \gamma_o} /2$ and $E = 4 \pm 3 \sqrt{ \gamma_e \gamma_o}/2$ each with degeneracy $2$.

In a similar, but much more tedious procedure, the eigenvalues and eigenvectors of the $N=8$, $256 \times 256$ $H$ matrix can be found explicitly.  The results for that case and the smaller-$N$ cases support the following claims, to be proven for general $N$ in the following section:
\begin{itemize}
   \item there are exactly two $0$ eigenvalue states; one in the even and one in the odd spin-parity subspace;
   \item the maximum eigenvalue in both the even and odd spin-parity blocks has value $N$;
   \item each $0$ eigenvalue state is in the $q=0$ subspace, with eigenvectors containing symmetric sums of shifted states;
   \item eigenvalues depend only upon the single parameter $\sqrt{\gamma_e \gamma_o}$.  This implies that the spectrum can be fully deduced through knowledge of the spectrum for the single-temperature case, $T_e = T_o$.
\end{itemize}

 \section{General case}

Here we approach the problem using standard methodology \cite{grynberg}.  Starting with the full, two-temperature operator in terms of spin raising and lowering operators, Eq. \ref{Hop},
we can rewrite in terms of fermionic operators ($c_{j}$, $c_{j}^{\dagger}$) that satisfy anti-commutation relations
\begin{eqnarray*}
\{c_{j},c_{i}^{\dagger}\} \equiv c_j c_i^{\dagger} + c_i^{\dagger} c_j =\delta_{j,i} \\
\{c_{j},c_{i}\}=\{c_{j}^{\dagger},c_{i}^{\dagger}\}=0 .
\end{eqnarray*}
This transformation, due to Jordan and Wigner \cite{jw} is as follows
\begin{eqnarray*}
\sigma_{j}^{+}&=& c_{j}^{\dagger}e^{i \pi \sum_{i<j}c_{i}^{\dagger}c_{i}}\\
\sigma_{j}^{-}&=& c_{j}e^{-i \pi \sum_{i<j}c_{i}^{\dagger}c_{i}} .
\end{eqnarray*}

Straightforward application of this transformation produces the 'fermionized' pseudo-Hamiltonian 
\begin{eqnarray} 
H&=&-\frac{1}{2} \sum_{j \; even}[c^{\dagger}_{j}c_{j+1}+c^{\dagger}_{j+1}c_{j}+(1+\gamma_e)c^{\dagger}_{j}c^{\dagger}_{j+1} - (1-\gamma_e)c_{j}c_{j+1} +\gamma_e (c^{\dagger}_{j}c_{j}+c^{\dagger}_{j+1}c_{j+1}) \nonumber \\
 &-&(1+\gamma_e)] -\frac{1}{2} \sum_{j \; odd}[c^{\dagger}_{j}c_{j+1}+c^{\dagger}_{j+1}c_{j}+(1+\gamma_o)c^{\dagger}_{j}c^{\dagger}_{j+1}- (1-\gamma_o)c_{j}c_{j+1} \nonumber \\
 &+& \gamma_o (c^{\dagger}_{j}c_{j}+c^{\dagger}_{j+1}c_{j+1})-(1+\gamma_o)] . \label{fermH} 
\end{eqnarray}

For the even spin-parity subspace the requirement that $\sigma^+_{N+1} = \sigma^+_1$ and $\sigma^-_{N+1} = \sigma^-_1$ implies $c_{N+1} = -c_1$ and $c^{\dagger}_{N+1} = -c^{\dagger}_1$.  For the odd spin-parity case,  $c_{N+1} = c_1$ and $c^{\dagger}_{N+1} = c^{\dagger}_1$.
We take advantage of the translation symmetry mentioned in Section 3 by defining two kinds of fermions in momentum space, one created from even-numbered cells, the other from odd-numbered cells.  Define the following momentum-space operators:
\begin{eqnarray*}
a_{q}^{\dagger} = e^{\frac{i\pi}{4}} {\sqrt{\frac2N}}\sum_{m \; even}c_{j}^{\dagger} e^{i \frac{j}{2} q  } & \mbox{        } &
c_{j}^{\dagger} = e^{\frac{-i\pi}{4}} {\sqrt{\frac2N}}\sum_{q \in Q}a_{q}^{\dagger} e^{-i \frac{j}{2} q  } \\ 
b_{q}^{\dagger} = e^{\frac{i\pi}{4}} {\sqrt{\frac2N}}\sum_{j \; odd}c_{j}^{\dagger} e^{i \frac{j+1}{2} q  } & \mbox{        } &
c_{j}^{\dagger} = e^{\frac{-i\pi}{4}} {\sqrt{\frac2N}}\sum_{q \in Q}b_{q}^{\dagger} e^{-i \frac{j+1}{2} q  } .
\end{eqnarray*}
We chose q values to belong to the set $Q = \{\pm \frac{2\pi}{N}, \pm \frac{6\pi}{N}, \pm \frac{10\pi}{N},\ldots \pm \frac{(N-2)\pi}{N} \}$ for states with even spin-parity, and $Q = \{0,\pm \frac{4\pi}{N}, \pm \frac{8\pi}{N}, \ldots \pm \frac{(N-4)\pi}{N}, \pi \}$ for states with odd spin-parity, to assure proper periodic boundary conditions for each case.  These definitions of $Q$ assume that $N/4$ is integer-valued, but the ultimate results are valid as long as $N$ is even.

Like the operators $c_{j}$ and $c_{j}^{\dagger}$, the momentum space operators obey the canonical fermionic anticommutation relationships: 
\begin{eqnarray*}
\{a_{q},a_{q'}\}&=&\{a_{q}^{\dagger},a_{q'}^{\dagger}\}=0 \mbox{\hspace{.5 in}}  \{a_{q},a_{q'}^{\dagger}\}=\delta_{q,q'} \\
\{b_{q},b_{q'}\}&=&\{b_{q}^{\dagger},b_{q'}^{\dagger}\}=0 \mbox{\hspace{.5 in}}  \{b_{q},b_{q'}^{\dagger}\}=\delta_{q,q'} \\
\{a_{q},b_{q'}\}&=&\{a_{q}^{\dagger},b_{q'}^{\dagger}\}=0 \mbox{\hspace{.5 in}}  \{a_{q},b_{q'}^{\dagger}\}=0 
\end{eqnarray*}
In terms of these operators, the pseudo-Hamiltonian is written as
\begin{eqnarray} H=\sum_{q \in Q}&&\left[-\cos(\frac{q}{2}) (e^{i\frac{q}{2}} a^{\dagger}_{q}b_{q}+ e^{-i\frac{q}{2}} b^{\dagger}_{q}a_{q}) \right. \nonumber \\
   &+& a^{\dagger}_{q}b^{\dagger}_{-q} \, \frac{i}{2}[(1+\gamma_e)e^{iq} - (1+\gamma_o)] + a_{q}b_{-q} \;\frac{i}{2}[(1-\gamma_e)e^{-iq} - (1-\gamma_o)]  \nonumber \\
    &-& \left. \frac{(\gamma_e+\gamma_o)}{2}(a^{\dagger}_{q}a_{q}+b^{\dagger}_{q}b_{q}) + 2+\gamma_e+\gamma_o   \right] .\label{H1} 
\end{eqnarray}

The ultimate step in the derivation is to perform a Bogoliubov type similarity transform to new variables in which $H$ takes diagonal form.  We postulate a diagonal form for $H$ that reads
\begin{equation} 
     H = \sum_q ( \omega_q \tilde{\chi_q} \chi_q + \omega'_q \tilde{\xi_q} \xi_q + \mbox{const.} ), \label{L1} 
\end{equation}
while assuming that the operators $\chi_q$ and $\xi_q$ obey Fermionic anticommutation relations.  Since $H$ is not Hermitian, $\tilde{\chi_q} \neq \chi_q^{\dagger}$, and $\tilde{\xi_q} \neq \xi_q^{\dagger}$.
It is easily demonstrated that
\begin{equation} 
     [\chi_q,H]_{-} =  \omega_q \chi_q \;\;\;\mbox{and}\;\;\; [\xi_q,H]_{-} =  \omega'_q \xi_q .  \label{L2} 
\end{equation}

Following Lieb, Schultz and Mattis\cite{lsm} we define the $\chi_q$ and $\xi_q$ variables in terms of the $a_q$ and $b_q$ variables in the following manner:
\begin{equation} 
     \chi_q = c_1 a^{\dagger}_q + c_2 b_{-q} + c_3 b^{\dagger}_q + c_4 a_{-q}      \label{L3} 
\end{equation}
and
\begin{equation} 
     \xi_q = d_1 a^{\dagger}_q + d_2 b_{-q} + d_3 b^{\dagger}_q + d_4 a_{-q}   ,   \label{L4} 
\end{equation}
where the $c_i$'s and $d_i$'s are constants.  We can calculate the commutator with $H$ in the form given by Eq. \ref{H1} of each term in $\chi_q$ as expressed in Eq. \ref{L3}, using
 \[   [a^{\dagger}_q,H]_{-} = \gamma_{av} a^{\dagger}_q + \epsilon ' b_{-q} + \cos{\frac{q}{2}}e^{-Iq/2} \;b^{\dagger}_q \]
 \[   [a_{-q},H]_{-} = -\gamma_{av} a_{-q} + \overline{\epsilon} b^{\dagger}_{q} - \cos{\frac{q}{2}}e^{-Iq/2}\; b_{-q} \]
 \[   [b^{\dagger}_q,H]_{-} = \gamma_{av} b^{\dagger}_q - \overline{\epsilon'} a_{-q} + \cos{\frac{q}{2}}e^{Iq/2}\; a^{\dagger}_q \]
 \[   [b_{-q},H]_{-} = -\gamma_{av} b_{-q} - \epsilon a^{\dagger}_{q} - \cos{\frac{q}{2}}e^{Iq/2}\; a_{-q} . \]
Constants in these expressions are defined as:
\[ \gamma_{av} \equiv \frac{\gamma_e+\gamma_o}{2} \;\;\;\; \gamma_{dif} \equiv \frac{\gamma_e-\gamma_o}{2} \]
\[ \epsilon \equiv  e^{iq/2} \left( -(1+\gamma_{av})\sin{\frac{q}2}+i \gamma_{dif} \cos{\frac{q}2} \right) \;\;\;\;
   \epsilon' \equiv  e^{-iq/2} \left( (1-\gamma_{av})\sin{\frac{q}2}-i \gamma_{dif} \cos{\frac{q}2} \right). \]
$\overline{\epsilon}$ represents $\epsilon$ as defined above but with $q$ replaced by $-q$, and likewise for $\epsilon'$.
Eq. \ref{L2}, with these results used to evaluate the left hand side, produces a large operator equation.  The coefficients of each of the four operators on the right-hand-side must equal the corresponding coefficient on the left-hand-side.  This requirement produces a linear set of equations for the constants $c_i$ that can be expressed by the matrix equation  $\omega_q C = M C$ where
\[   C = \left( \begin{array}{c} c_i \\ c_2 \\ c_3 \\ c_4 \end{array}  \right) \]
and
\begin{equation}
   M = 
 \left[ \begin {array}{cccc} \gamma_{av} & -\epsilon & \cos{\frac{q}2} e^{Iq/2} & 0 \\ 
                             \epsilon' & -\gamma_{av} & 0 & -\cos{\frac{q}2} e^{-Iq/2} \\
                             \cos{\frac{q}2} e^{-Iq/2} & 0 & \gamma_{av} & \overline{\epsilon} \\
                             0 & -\cos{\frac{q}2} e^{Iq/2} & -\overline{\epsilon'} & -\gamma_{av} \end{array} \right] . \label{Mdef}
\end{equation}
 
Obviously, the eigenvalues of $M$ are possible values of the energies ($\omega_q$) of the $\chi$ excitations, which is what we seek.  Identical analysis of the $d_i$ coefficients show that the eigenvalues of the same matrix are possible values of the energies ($\omega'_q$) of the $\xi$ excitations.  The eigenvector components are quite complicated, but the eigenvalues simplify easily. Two of the eigenvalue are negative, thus inappropriate candidates for the excitation energies (that we know to be positive.)  The other two eigenvalues are $1 \pm \cos{\frac{q}{2}} \sqrt{\gamma_e \gamma_o}$.  Identify $ \omega_q = 1+\cos{\frac{q}{2}} \sqrt{\gamma_e \gamma_o}$ and $ \omega'_q = 1-\cos{\frac{q}{2}} \sqrt{\gamma_e \gamma_o}$ in order to deduce the exact spectrum for the model.

There is a unique even spin-parity vacuum state $\ket{0}_e$ defined by the relations:
\[  \chi_q \ket{0}_e = 0, \;\; \xi_q \ket{0}_e = 0 \;\;\mbox{for}\;\; q \in \{\pm \frac{2\pi}{N}, \pm \frac{6\pi}{N}, \pm \frac{10\pi}{N},\ldots \pm \frac{(N-2)\pi}{N} \}. \]  
Other even spin-parity states are formed by an even number of excitations of the $\tilde{\chi}_q$ or $\tilde{\xi}_q$ type:  each excitation of the former type carries energy $\omega_q$, and each of the latter type carries $\omega'_q$.  Since these excitations are fermionic, there cannot be two $\tilde{\chi}_q$ excitations with the same $q$, nor can there be two $\tilde{\xi}_q$ excitations with the same $q$.  The highest energy state has $N/2$ distinct $\tilde{\chi}_q$ excitations and $N/2$ distinct $\tilde{\xi}_q$ excitations, and carries total energy $N$.

The odd spin-parity sector of the spectrum also has a unique vacuum:
\[  \chi_q \ket{0}_o = 0, \;\; \xi_q \ket{0}_o = 0 \;\;\mbox{for}\;\-; q \in \{0,\pm \frac{4\pi}{N}, \pm \frac{8\pi}{N}, \ldots \pm \frac{(N-4)\pi}{N}, \pi \} . \]  
Other odd-parity states are formed by an even number of excitations of the $\tilde{\chi}_q$ or $\tilde{\xi}_q$ type with $q$ with values from the list appropriate to odd spin-parity.  The highest energy odd spin-parity state has $N/2$ distinct $\tilde{\chi}_q$ excitations and $N/2$ distinct $\tilde{\xi}_q$ excitations, and carries total energy $N$.

\section{Consequences and conclusions}

It is surprising that a relatively simple set of eigenvalues emerges from the great algebraic complexity of the solution for arbitrary $N$ of the two-temperature model considered herein.  The eigenvectors exhibited in Fig. 2  for the $N=4$ case show that even in this relatively simple case, the state vectors are algebraically difficult.   In particular, the $E=0$ eigenstate that corresponds to the steady-state solution, is not easily characterized.  In general the eigenstates depend separately upon the values of $\gamma_e$ and $\gamma_o$.  While the methodology employed in the previous section can, in principle, allow the extraction of the eigenvectors of $H$, the algebraic complexity of the eigenvectors of the matrix $M$ suggest that such a straightforward exposition of them is not likely to be illuminating.  Nonetheless, our results produce numerical expressions for the eigenstates given specific values for the two temperatures, and thus to expressions for particle densities and correlation functions.  General relationships for these have been previously exhibited by Mobilia \it et al.\rm  \cite{mobilia}

The fact that the eigenvalues depend upon the single parameter $\sqrt{\gamma_e \gamma_o}$ allows some simple deductions regarding special cases of the two-temperature model.  If one of the temperature baths has infinite temperature (e.g. $\gamma_{e}=0$), the eigenvalues are those of the case of a single-temperature model with both baths at infinite temperature.  These eigenvalue are identical to those of the Glauber model \cite{glauber}, but the eigenvectors are of greater complexity.  If one of the temperature baths has $T=0$ (e.g. $\gamma_{e}=1$ ), the energy spectrum becomes the same as the for the one-temperature case related to the temperature of the other bath.  In the RDS language, this case corresponds to a system with pair creation prohibited, and pair annihilation at a rate of $1$ for the even sites. In general, for every case with distinct temperatures $T_e$ and $T_o$, there is a single temperature that will yield the same energy eigenvalues. The spectrum of energies for the single temperature case follows from the work of Grynberg \it et al.\rm \cite{grynberg} (for the special case $h=h$'$=\frac12$, $\epsilon = (1+\gamma)/2$, $\epsilon$'$= (1-\gamma)/2$ with $\gamma = \tanh{(2/k_B T)}$. 

A few general observations about the spectrum in the thermodynamic limit, $N \rightarrow \infty$, are possible.  Paralleling an observation of Grynberg, \it et. al. \rm \cite{grynberg} for the single temperature case, as long as $\sqrt{\gamma_e \gamma_o} < 1$ (at least one temperature bath is above absolute zero), there is a gap between the ground state and the next-highest energy level of $2(1-\sqrt{\gamma_e \gamma_o})$.  This assures that states other than the steady state decay exponentially in time.  The spectrum of remaining states consist of bands of energy levels centering on states with $E=4,6,8,\ldots$.  The widths of these bands grows with $E$, while the spacing between adjacent states remains constant at $2$.  As a result, regardless of how small the parameter $\sqrt{\gamma_e \gamma_o}$ is, the bands will overlap for high energies.  These observations are separately true for the even spin-parity and odd spin-parity segments of the energy spectrum.

Because $H$ does not cause transitions between states of different spin-parity, the time evolution of any initial state can be broken into two independent segments.  Any initial configuration of spin states can be broken into a piece with even spin-parity with probability $P_e$ and one of odd spin-parity with probability $P_o$, with $P_e+P_o=1$.  The even spin-parity segment decays towards the steady state $\ket{0}_e$ maintaining constant probability $P_e$; likewise the odd segment decays towards  $\ket{0}_o$ maintaining constant probability $P_o$.  The separate even and odd spin-parity energies control the rate of decay of the non-steady-state components for each segment.

Symmetries discussed in Section 2 enable a few additional comments about the general form of the eigenvectors.  Because we know the eigenvalues to be invariant under the interchange $\gamma_e \leftrightarrow \gamma_o$ it follows that displacement of the ring by a single site (effectively a permutation of basis states) along with an interchange of values of $\gamma_e$ and $\gamma_o$ should transform any eigenvector into a constant multiple of itself, or into a different eigenvector with the same eigenvalue.  Since the ground state ($E=0$) and the maximum energy state ($E=N$) for the even spin-parity sector and for the odd spin-parity sector are nondegenerate, each should be invariant within a constant under this transformation. 

The eigenvalues are also invariant under the simultaneous change of sign of $\gamma_e$ and $\gamma_o$.  This implies that a simultaneous flip of all spins (another basis state permutation) accompanied by a sign change in both $\gamma$'s should transform an eigenvector into a constant multiple of itself or into another eigenvector with the same $E$.  The non-degenerate eigenstates in each spin-parity sector should transform into constant multiples of themselves under this symmetry transformation. 

Mobilia \it et al. \rm \cite{mobilia} have examined the behavior of this model in the case where the $\gamma$'s have opposite signs.  Although the concept of negative temperatures does not make physical sense for the KISC model, in the context of the RDS system it corresponds to a grid where the pair creation rate exceeds the pair annihilation rate on one sub-lattice, and the opposite is true for the other sub-lattice.  The eigenvalues for this case will have positive real parts ($Re(E) = 0,2,4, \ldots, N$) corresponding to exponential damping in time, and imaginary parts proportional to $\sqrt{|\gamma_e \gamma_o|}$ producing oscillatory behavior.  This result is consistent with the predictions presented in Mobilia \it et al. \rm \cite{mobilia}, for example that under these conditions the density of particles approaches its equilibrium value via a term proportional to $\exp(-2t)\sin(2\sqrt{|\gamma_e \gamma_o|}t + \delta)$.  Our spectrum results show such a behavior for the highest frequency oscillation associated with the most-slowly decaying component of the state function evolving from a general initial condition.

In ongoing work we seek a compact expression for the steady state of these models. Given the relationship between spin systems and reaction-diffusion systems, it will be interesting to investigate the effect of various initial conditions and open boundary conditions on the dynamics of the system. From an experimental point of view, open boundary conditions for RDS systems would be important in the study of chemical reactions that include creation and annihilation processes and dimer deposition. Although particle densities and correlation functions can be calculated fairly straightforward in the thermodynamic limit, the finite-size effects may also be worth investigating.

We can also imagine other extensions of the models presented. For example, we are interested in considering an RDS model with non-uniform diffusion rates for the odd and even sites, and different creation and annihilation rates. This can also shed some light on the general problem of dimerized spin chains \cite{giorgi}.

\section{Acknowledgments}
I. M. wants to express special thanks to KITP for hospitality and financial support.This research was supported in part by the National Science Foundation under
Grant No. PHY05-51164.

\begin{figure}[] % float placement: (h)ere, page (t)op, page (b)ottom, other (p)age
  \centering
  % file name: C:/Winter09/Kinetic Ising/KISC_mapping.eps
  \includegraphics[bb=15 13 597 780,width=5.67in,height=7.47in,keepaspectratio]{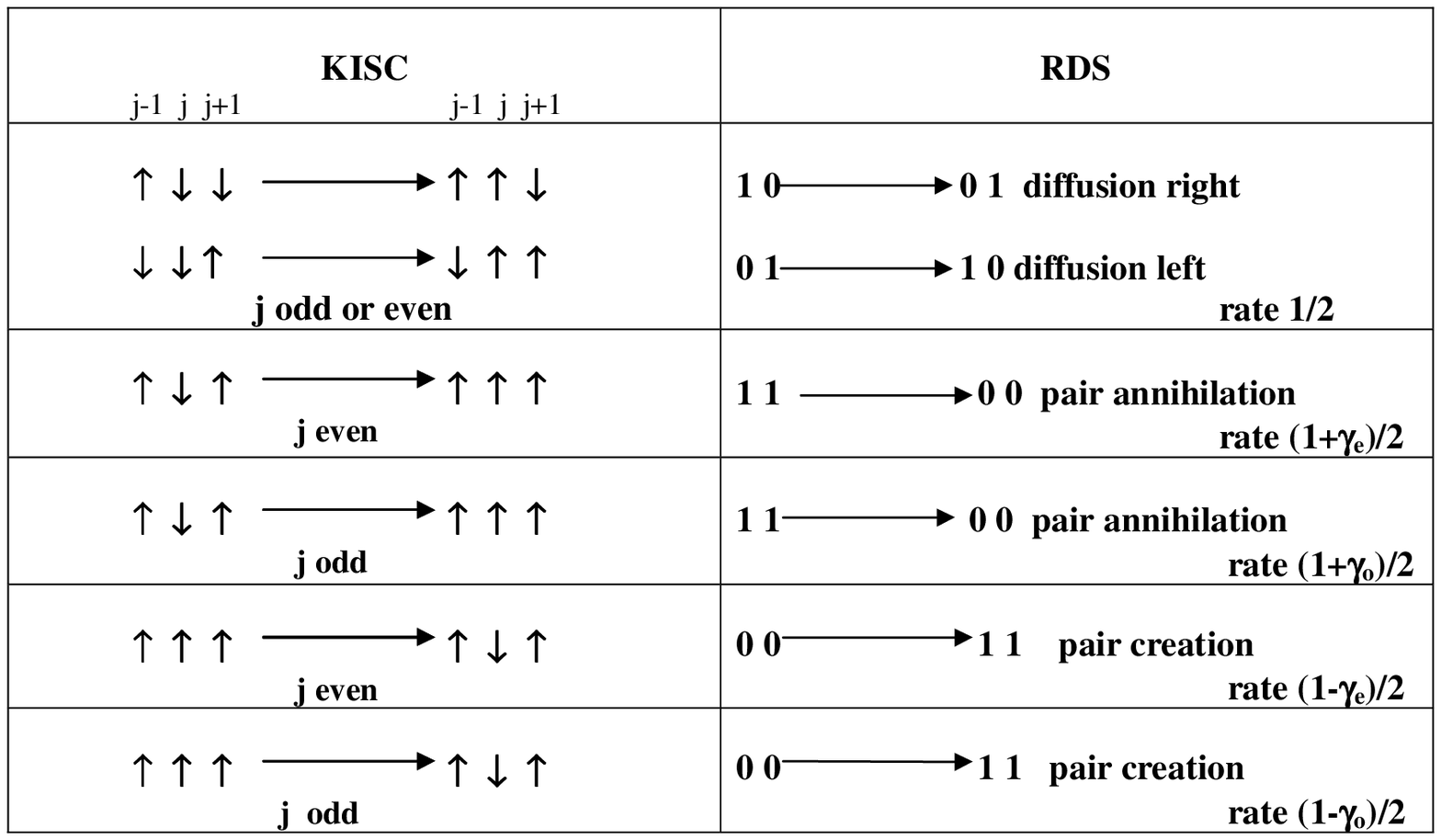}
  \caption{Correspondence between the two temperature kinetic Ising chain (KISC) and the equivalent reaction-diffusion model (RDS.)}
  \label{fig:KISC_mapping}
\end{figure}

%\begin{figure}[] % float placement: (h)ere, page (t)op, page (b)ottom, other (p)age
%  \centering
%  % file name: C:/Winter09/Kinetic Ising/table_eqG.eps
%  \includegraphics[bb=15 13 597 780,width=5.67in,height=7.47in,keepaspectratio]{table_eqG}
%  \caption{Caption for table\_eqG}
%  \label{fig:table_eqG}
%\end{figure}

%\begin{figure}[] % float placement: (h)ere, page (t)op, page (b)ottom, other (p)age
%  \centering
%  % file name: C:/Winter09/Kinetic Ising/table_eqG1.eps
%  \includegraphics[bb=15 13 597 780,width=5.67in,height=7.47in,keepaspectratio]{table_eqG1}
%  \caption{Caption for table\_eqG1}
%  \label{fig:table_eqG1}
%\end{figure}

\begin{figure}[] % float placement: (h)ere, page (t)op, page (b)ottom, other (p)age
  \centering
  % file name: C:/Winter09/Kinetic Ising/table_neqG.eps
  \includegraphics[bb=15 13 597 780,width=5.67in,height=7.47in,keepaspectratio]{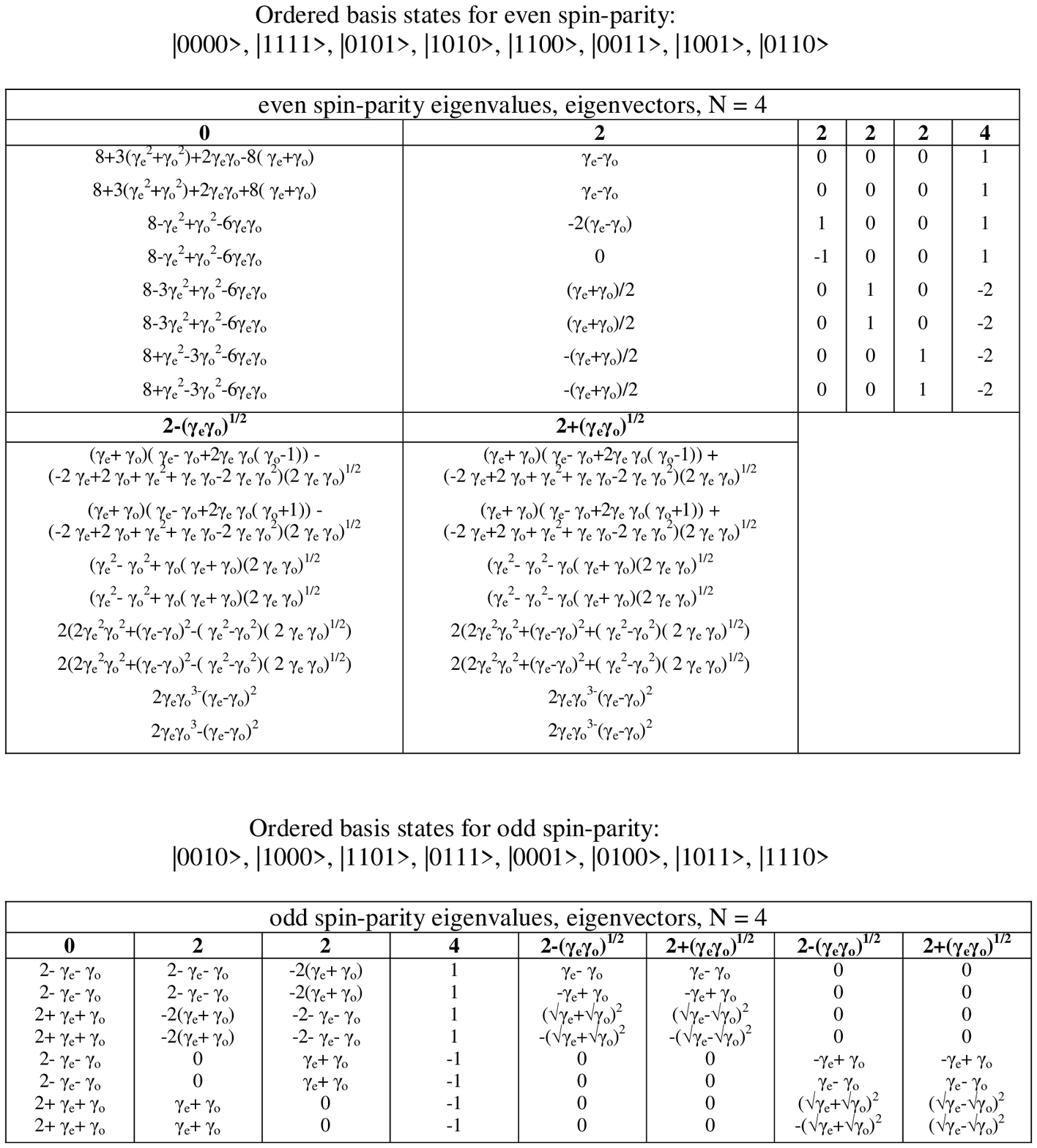}
  \caption{Eigenstates (un-normalized) for the $N=4$ case of the two-temperature kinetic Ising model.}
  \label{fig:Neq4}
\end{figure}

\end{document}